\documentclass[sigconf]{acmart}

\usepackage{enumitem}
\usepackage{mathtools}
\usepackage{multirow}
\usepackage{xcolor}
\usepackage{pifont}
\usepackage{url}
\usepackage{makecell}
\usepackage{stfloats}

\definecolor{framebg}{RGB}{255,255,255}
\definecolor{frameborder}{RGB}{0,0,0}

\newcommand{\mybox}[1]{%
 \vspace{1pt}
 \noindent
 \fcolorbox{frameborder}{framebg}{%
   \parbox{\dimexpr\linewidth-2\fboxsep-2\fboxrule}{%
     \small #1%
   }%
 }%
 \vspace{1pt}
}

\AtBeginDocument{%
  }

\copyrightyear{2024}
\acmYear{2024}
\setcopyright{rightsretained}
\acmConference[ICAIF '24]{5th ACM International Conference on AI in Finance}{November 14--17, 2024}{Brooklyn, NY, USA}
\acmBooktitle{5th ACM International Conference on AI in Finance (ICAIF '24), November 14--17, 2024, Brooklyn, NY, USA}
\acmDOI{10.1145/3677052.3698628}
\acmISBN{979-8-4007-1081-0/24/11}

\begin{document}

\title{A Dutch Financial Large Language Model}


\author{Sander Noels}
\orcid{0000-0001-6042-8461}
\affiliation{%
  \institution{Ghent University}
  \city{Ghent}
  \country{Belgium}}
\email{sander.noels@ugent.be}
\affiliation{%
  \institution{Silverfin}
  \city{Ghent}
  \country{Belgium}}
\email{sander.noels@silverfin.com}

\author{Jorne De Blaere}
\orcid{0009-0002-5911-9703}
\affiliation{%
  \institution{Silverfin}
  \city{Ghent}
  \country{Belgium}}
\email{jorne.deblaere@silverfin.com}

\author{Tijl De Bie}
\orcid{0000-0002-2692-7504}
\affiliation{%
  \institution{Ghent University}
  \city{Ghent}
  \country{Belgium}
}
\email{tijl.debie@ugent.be}

\renewcommand{\shortauthors}{Noels et al.}

\begin{abstract}
This paper presents FinGEITje, the first Dutch financial Large Language Model (LLM) specifically designed and optimized for various financial tasks. Together with the model, we release a specialized Dutch financial instruction tuning dataset with over 140,000 samples, constructed employing an automated translation and data processing method. The open-source data construction method is provided, facilitating the creation of financial instruction datasets in different languages. To evaluate model performance, the study introduces the first Dutch financial evaluation benchmark, along with an automated evaluation method that utilizes an LLM as an independent evaluator, reducing manual intervention in performance evaluation. The experimental results highlight the superior performance of FinGEITje across five critical Dutch and English financial tasks.
\end{abstract}




\maketitle

\section{Introduction}
\label{sec:introduction}

Recent advances in Natural Language Processing (NLP) have significantly transformed the way we analyze unstructured financial data \cite{araci2019finbert,wu2023bloomberggpt}. The advent of LLMs has further expanded the capabilities of NLP, demonstrating exceptional performance across various financial tasks. However, substantial challenges remain, particularly with non-English languages such as Dutch, which are underrepresented by current financial models \cite{nie2024survey}.

The highly technical nature of financial texts requires domain-specific LLMs to understand complex financial language and concepts effectively \cite{araci2019finbert, wu2023bloomberggpt}. The financial sector struggles to create these models due to the scarcity of annotated datasets and high data annotation costs \cite{wang2023fingpt}. In addition, existing financial LLMs mainly focus on English, creating a knowledge gap in localized financial contexts in languages such as Dutch. As a result, Dutch financial documents are often misinterpreted, causing inefficiencies and inaccuracies when using general models for Dutch financial tasks.

Despite approximately 24 million Dutch speakers worldwide, existing Dutch LLMs have not focused on financial documents. Tailored financial models have consistently shown superior performance in domain-specific tasks compared to generalized LLMs. This emphasizes the necessity of developing specialized models, such as an open-source, transparent, and accessible financial LLM for the Dutch language. Such a model could democratize financial data analysis and foster advances in both academia and industry \cite{wang2023fingpt,xie2023pixiu}.

\noindent Our main contributions are summarized as follows:
\begin{itemize}[nosep]
\item We introduce FinGEITje, the first open-source, transparent, and accessible Dutch financial LLM.
\item We provide the first extensive Dutch financial instruction tuning dataset and a comprehensive evaluation benchmark for Dutch financial LLMs.
\item We release a methodology for constructing financial instruction tuning datasets in multiple languages, supporting the development of financial LLMs across linguistic contexts.
\item We introduce an automated evaluation method using an LLM as an independent evaluator, enabling large-scale evaluations.
\end{itemize}

FinGEITje offers valuable applications for financial professionals and researchers, including sentiment classification of financial news and social media posts, key entity identification in financial documents, and news headline classification to validate claims about price movements. It also supports financial relation extraction and can answer specific financial queries, thus streamlining information retrieval for financial decision-making processes.

The remainder of this paper is as follows. Section \ref{sec:related-work} reviews the related work. Section \ref{sec:methodology} details our proposed model, FinGEITje, followed by an empirical evaluation in Section \ref{sec:experiments}. Section \ref{sec:conclusion} provides the conclusions of our study. Section \ref{sec:futurework-limitations} explores potential limitations and future research directions. Finally, Section \ref{sec:ethics} discusses the ethical considerations associated with our work.

\section{Related Work}
\label{sec:related-work}

To understand the context of our work, we review the literature on financial NLP models. We explore the challenges these models face, the necessity of financially-tailored LLMs, the scarcity of non-English language financial models, and the importance of Dutch-focused financial LLMs. Lastly, we consider the need for open-source and transparent financial LLMs to foster progress in this specialized field.

\subsection{Challenges of Financial NLP Models}

NLP has made significant strides in handling unstructured financial data, turning complex documents into clear insights and valuable market information. These advances have opened up various capabilities, ranging from stock movement prediction to more advanced financial tasks \cite{nie2024survey}. However, traditional NLP models face notable challenges in the financial sector.

One primary issue is the intrinsic complexity of financial language and jargon, which often results in a gap in understanding domain-specific documents \cite{araci2019finbert,wu2023bloomberggpt,noels2024topo}. This complexity requires sophisticated models capable of accurately interpreting specialized terminology. Another significant barrier is the scarcity of annotated datasets, exacerbated by the high cost associated with data annotation in this specialized field \cite{wang2023fingpt}. A third challenge lies in the limited inferential capabilities of traditional models, particularly in critical tasks such as making informed investment decisions \cite{wang2023fingpt}. Lastly, the widespread adaptability of many NLP models remains constrained, as they are often optimized for a single task and lack the ability to generalize across multiple tasks \cite{mishra2021cross}.

Due to these challenges, it is essential to develop more advanced, versatile, and robust NLP models tailored to the dynamic and complex requirements for understanding financial documents. In this evolving landscape, the need for financially tailored LLMs has become increasingly evident.

\subsection{The Success of Financially-tailored LLMs}

The financial industry has recently seen a surge in the application of LLMs due to their remarkable capabilities in performing various tasks using natural language instructions, which often do not require extensive retraining on specific data \cite{la2020end,deng2023llms,noels2023automated, brown2020language,sanh2021multitask}. Unlike traditional financial NLP models that typically excel only in narrowly defined tasks, LLMs exhibit substantial versatility across various financial applications.

This progress witnessed a significant boost with FinBERT, a domain-specific adaptation of BERT, showing that fine-tuning on financial documents greatly increases performance in financial tasks \cite{araci2019finbert}. The success of FinBERT spurred a series of innovative models, notably BloombergGPT, FinGPT, FinMA, and FinTral, each advancing the integration of financial data and improving model performance \cite{wu2023bloomberggpt,yang2023fingpt,xie2023pixiu,bhatia2024fintral}.

Earlier models were limited by their relatively small parameter sizes, typically below one billion parameters, which constrained their generalization capabilities. The push towards multi-billion parameter models has significantly improved their ability to generalize and perform across various financial tasks \cite{xie2023pixiu}. This scalability plays a crucial role in the effectiveness of models in handling complex financial documents and generating precise and insightful outputs.

\subsection{The Scarcity of Non-English Financial LLMs}

One major issue in the current state of financial LLMs is the lack of models for languages other than English. Because English is a globally prominent language for international finance as well as machine learning research, it has been the focus for the majority of the existing work on financial LLMs. However, the comparative lack of models available in other languages represents a significant knowledge gap, preventing an in-depth understanding of localized financial contexts and nuances. The development of LLMs in these other languages is dependent on the availability of sufficient datasets and appropriate evaluation methods. Although extensive datasets are available for English, there is a notable scarcity of such datasets for other languages \cite{nie2024survey}.

The development of financial LLMs in languages other than English is still at an early stage. Notable examples include the Mengzi-fin \cite{zhang2021mengzi} and BBT-FinT5 \cite{lu2023bbt} models, both uniquely designed for the Chinese language. In addition to non-English language-specific models, multilingual financial LLMs, such as FinMA-ES \cite{zhang2024d} and ICE-PIXIU \cite{hu2024no}, have been developed, which have been fine-tuned on Spanish and English data and Chinese and English data, respectively. These models attempt to capture the linguistic nuances of the financial domain in various languages.

Language-specific financial benchmarks are a crucial factor for evaluating the impact and effectiveness of these financial LLMs in diversified linguistic contexts \cite{nie2024survey}. The gap between English and other languages in the domain of financial language models is the absence of financial benchmarks tailored to those languages. A worthy mention is a newly established benchmark for Japanese financial LLMs, although it does not accompany a specified LLM \cite{hirano2024construction}.

Table \ref{tab:models-comparison} compares the different financial LLMs available today across several parameters, including language compatibility. The addition of Chinese-specific models and multilingual models indicate a diversification in language applicability beyond English. For a complete analysis of the existing financial LLMs, see \cite{nie2024survey}.

\begin{table*}[htbp]
\footnotesize
\centering
\begin{tabular}{p{2.5cm}lccccccc}
\toprule
\textbf{Model} & \textbf{Backbone} & \textbf{Size} & \multicolumn{2}{c}{\textbf{Open Source}} & \textbf{Instruct} & \textbf{Language} & \textbf{Release Date} \\
\cmidrule(lr){4-5} 
 & & & \textbf{Model} & \textbf{Data} & & & &\\
\midrule
FinBERT \cite{araci2019finbert} & BERT & 110M & \ding{51} & \ding{51} & \ding{55} & English & 08/2019 \\
Mengzi-fin \cite{zhang2021mengzi} & RoBERTa & 103M & \ding{51} & \ding{55} & \ding{55} & Chinese & 10/2021 \\
BBT-FinT5 \cite{lu2023bbt} & T5 & 220M & \ding{51} & \ding{51} & \ding{55} & Chinese & 02/2023 \\
BloombergGPT \cite{wu2023bloomberggpt} & BLOOM & 50B & \ding{55} & \ding{55} & \ding{55} & English & 03/2023 \\
FinGPT \cite{wang2023fingpt} & LLaMA & 7/13B & \ding{51} & \ding{51} & \ding{51} & English & 10/2023 \\
FinMA \cite{xie2023pixiu} & LLaMA & 7/30B & \ding{51} & \ding{51} & \ding{51} & English & 06/2023 \\
FinMA-ES \cite{zhang2024d} & LLaMA2 & 7B & \ding{51} & \ding{51} & \ding{51} & English, Spanish & 02/2024 \\
FinTral \cite{zhang2024d} & Mistral & 7B & \ding{55} & \ding{55} & \ding{51} & English & 02/2024 \\
ICE-PIXIU \cite{hu2024no} & LLaMA2 & 7B & \ding{51} & \ding{51} & \ding{51} & English, Chinese & 03/2024 \\
\bottomrule
\end{tabular}
\caption{\label{tab:models-comparison}Comparison of pre-existing LLMs in the financial domain.}
\end{table*}

\subsection{The Importance of Dutch-focused Financial LLMs}

The implementation of financial LLMs is largely centered around a few global languages, mainly English and Chinese, with little attention to languages like Dutch. This lacks a sufficient focus on understanding financial contexts where the Dutch language is dominant. An estimated 24 million people converse in Dutch around the world according to the Dutch \textit{Taalunie} (Language Union), suggesting a considerable demand in developing a Dutch-specific financial LLM.\footnote{\url{https://taalunie.org/informatie/24/feiten-cijfers}}

Generalized models, particularly those predominantly trained on English data, face a significant obstacle in their inability to generate Dutch sentences. They often revert to English after a few sentences or provide inaccurate translations from English to Dutch \cite{vanroy2023language}. This tendency leads to a loss of the finer nuances and subtleties inherent in the Dutch language. For a Dutch-specific LLM to retain its credibility and applicability, it is important to preserve linguistic subtleties and maintain the natural flow of the language, while avoiding the interweaving of English terms.

Moreover, the scarcity of Dutch-centered LLMs is not only due to the deficiency of pre-established, pre-trained models. It is also a reflection of the significant gap in critical supporting infrastructure, which includes specialized datasets, benchmarks, and leaderboards \cite{vanroy2023language}. In an attempt to address this gap, several Dutch pre-trained models have been introduced, such as \texttt{GEITje-7B}\footnote{\url{https://huggingface.co/Rijgersberg/GEITje-7B}}, \texttt{GEITje-7B-ultra-sft}\footnote{\url{https://huggingface.co/Rijgersberg/BramVanroy/GEITje-7B-ultra-sft}}, \texttt{GEITje-7B-ultra}\footnote{\url{https://huggingface.co/BramVanroy/GEITje-7B-ultra}}, \texttt{fietje-2b}\footnote{\url{https://huggingface.co/spaces/BramVanroy/fietje-2b}}, and \\ \texttt{Reynaerde-7B-Chat}\footnote{\url{https://huggingface.co/ReBatch/Reynaerde-7B-Chat}}. However, their focus does not specifically fall on Dutch financial documents, thus limiting their effectiveness in financial applications.

In summary, the need for LLMs proficient in Dutch financial contexts cannot be overstated. Developing Dutch-centric financial LLMs is crucial for generating insightful and accurate interpretations from financial documents, capturing the richness and specificity of Dutch financial language.

\subsection{The Need for Open-Source Financial LLMs and Their Transparency}

The rise of proprietary financial LLMs, such as BloombergGPT \cite{wu2023bloomberggpt}, have utilized their exclusive access to specialized and nuanced financial data, leading to a pressing need for greater accessibility, transparency, and democratization of financial data \cite{araci2019finbert}. 

Open-source alternatives, such as FinGPT \cite{wang2023fingpt}, have shown the value of democratization in this domain, demonstrating similar or even superior performance to proprietary models with a lower investment in resources \cite{yang2023fingpt}. However, the progress made in creating open-source financial LLMs, which primarily focus on English, has overlooked the development of Dutch or other non-English specific financial models.

The creation of a larger suite of open-source Dutch LLMs will accelerate both academic and industrial research in automated language generation, empowering non-expert users to employ generative AI tools in their native language. The open-source approach creates an honest culture, supports fair practices, and allows for clear checks on how the model works and behaves.

\section{Methodology}
\label{sec:methodology}

In this section, we provide a detailed explanation of the methodology adopted for constructing FinGEITje, the first Dutch financial LLM. We begin by outlining the dataset preparation, proceed with the instruction tuning process, and end with a description of our Dutch financial evaluation benchmark.

\subsection{Dataset}
\label{subsec:dataset}

The dataset forms the backbone of our financial LLM, FinGEITje. Here, we detail the process of collecting raw data, constructing instruction datasets, and translating them into Dutch. We also discuss deduplication and filtering processes to maintain data quality.

\subsubsection{Raw Data}

The development of a high-quality Dutch financial LLM requires an effective dataset. We constructed an instruction tuning dataset specifically for performing Dutch financial tasks. Our dataset creation methodology draws much of its inspiration from FinGPT \cite{wang2023fingpt}. Following FinGPT, we mainly rely on open-source data instead of depending on the self-instruct method \cite{wang2022self}, such as Alpaca, due to several advantages: high-quality annotations by domain experts, minimal costs, no restrictions on commercial use, and a diverse range of text types and modalities. These sources include news articles, financial reports, tweets, and multi-modal data such as time series data and tables. We transform the English datasets into a Dutch instruction tuning dataset using an automated translation and data processing method, explained in Section \ref{sec:translate}.

\begin{table*}[htbp]
\footnotesize
\centering
\begin{tabular}{p{2cm}p{2.2cm}p{1.2cm}p{1.2cm}p{2cm}p{1.5cm}p{3cm}}
\toprule
\textbf{Data} & \textbf{Task} & \textbf{Raw Data} & \textbf{Instruction Data} & \textbf{Data Types} & \textbf{Modalities} & \textbf{Source}\\ 
\midrule
FPB \cite{malo2014good} & sentiment analysis & 4,845 & 4,845 & news & text & \texttt{FinGPT/ fingpt-sentiment-train} \\ 
FiQA-SA \cite{maia201818} & sentiment analysis & 1,213 & 1,213 & news headlines, tweets & text & \texttt{FinGPT/ fingpt-sentiment-train} \\ 
TFNS \cite{magic2022twitter} & sentiment analysis & 9,543 & 9,543 & tweets & text & \texttt{FinGPT/ fingpt-sentiment-train} \\ 
NWGI & sentiment analysis & 16,184 & 16,184 & news & text & \texttt{FinGPT/ fingpt-sentiment-train} \\ 
NER \cite{alvarado2015domain} & named entity recognition & 609 & 609 & financial agreements & text & \texttt{FinGPT/fingpt-ner} \\ 
NER (CLS) \cite{alvarado2015domain} & named entity recognition & 609 & 17,051 & financial agreements & text & \texttt{FinGPT/fingpt-ner-cls} \\ 
Headline \cite{sinha2021impact} & news headline classification & 11,412 & 102,708 & news headlines & text & \texttt{FinGPT/fingpt-headline} \\ 
FinRed \cite{sharma2022finred} & relationship extraction & 6,768 & 32,670 & news, earning call transcripts & text & \texttt{FinGPT/fingpt-finred} \\ 
Finance Alpaca & question answering & 68,912 & 68,912 & news headlines, tweets, earnings reports & text & \texttt{gbharti/finance-alpaca} \\  
FiQA-QA \cite{maia201818} & question answering & 17,110 & 17,110 & earnings reports & text, table & \texttt{FinGPT/fingpt-fiqa\_qa} \\ 
ConvFinQA \cite{chen2022convfinqa} & question answering & 12,594 & 12,594 & earnings reports & text, table & \texttt{FinGPT/fingpt-convfinqa} \\ 
\bottomrule
\end{tabular}
\caption{\label{tab:datasets}Summary of datasets used for training the Dutch financial LLM.}
\end{table*}

Table \ref{tab:datasets} provides an overview of the key datasets used, encompassing various financial tasks:

\textbf{Sentiment analysis} evaluates the sentiment of financial texts such as news articles and tweets, classifying them as positive, negative, or neutral. \textbf{Named entity recognition} identifies and classifies key financial entities, such as persons, organizations, and locations, facilitating the construction of financial knowledge graphs. \textbf{News headline classification} examines financial headlines, determining the accuracy of statements about price movements. \textbf{Relation extraction} seeks to identify and extract various financial relationships present within textual data, while the \textbf{question answering} task automatically provides answers to financial questions based on text-based information. For more details on our dataset construction, please refer to our GitHub repository.\footnote{\url{https://github.com/snoels/fingeit}}

\subsubsection{Instruction Dataset Construction}
\label{subsubsec:dataset-construction}

The construction of an English financial instruction tuning dataset serves as the starting point in this process, aligning each raw data sample with one or more instruction prompts relevant to specific financial tasks. The creation of additional instruction prompts is crucial as it effectively increases the overall number of data samples \cite{yang2023fingpt}.

\begin{figure}[htbp]
\centering
\includegraphics[width=\columnwidth]{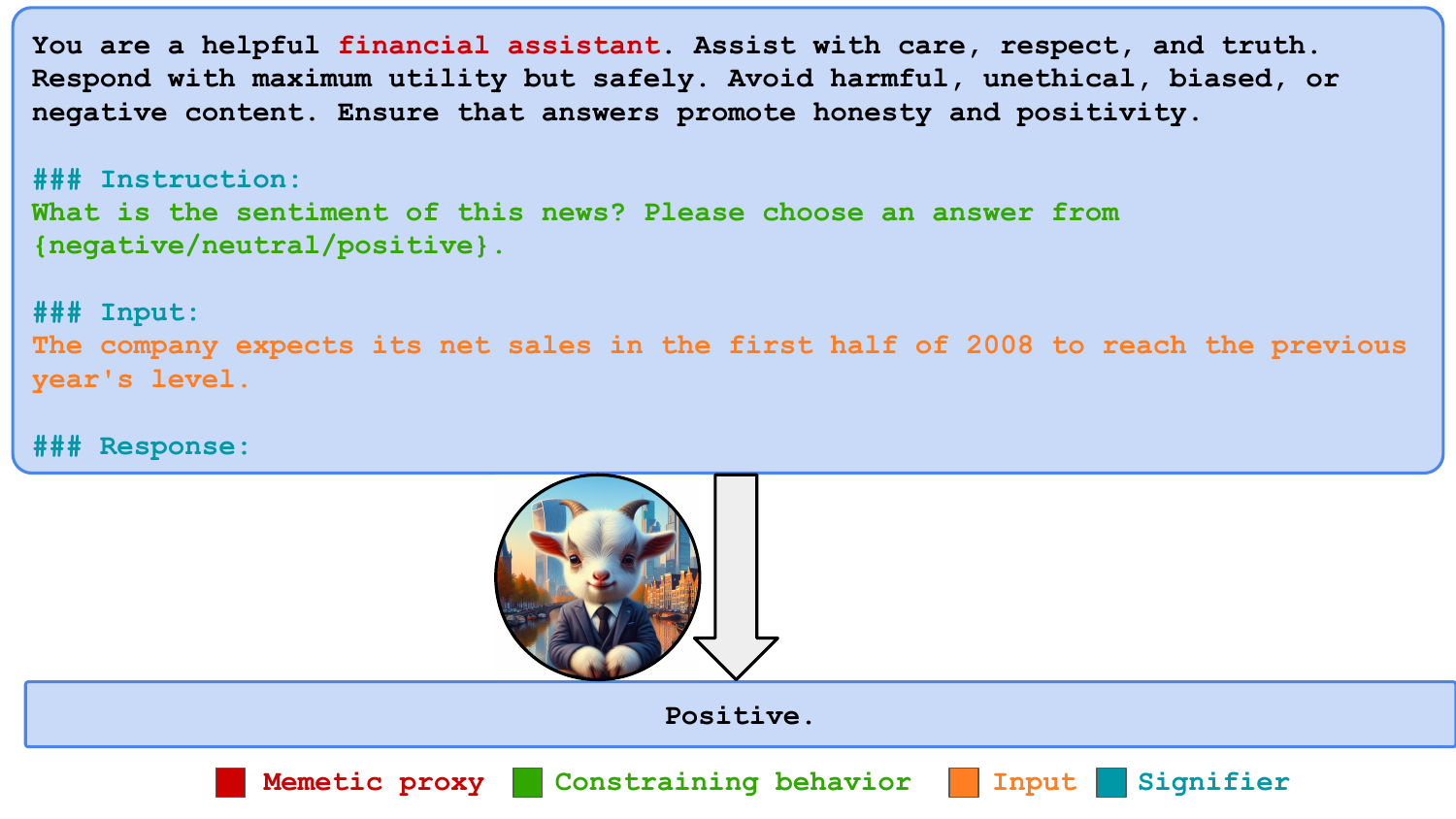}
\caption{The prompting method of FinGEITje.}
\label{fig:prompt}
\end{figure}

Our prompt construction is based on an innovative method tailored to the needs of financial LLMs. An illustration of how our prompting method operates is depicted in Figure \ref{fig:prompt}. In our methodology, we attribute a memetic proxy \cite{reynolds2021prompt} to the model, simulating the persona of a financial expert. Following the memetic proxy, we present the task-linked instruction prompts, which further constrain the model to refine its output. The input provides additional financial context, and the response completes the instruction output.

The proper formulation of these instructions is pivotal in instruction fine-tuning, as they act as guides in training the model to perform specific tasks. From FinGPT, we adopt the instruction formulations for all tasks \cite{yang2023fingpt}, except Finance Alpaca, which derives instructions based on the self-instruct method \cite{wang2022self}. Table \ref{tab:datasets} provides information on how many instruction tuning samples are generated for each dataset. For a more detailed explanation of how the English financial instruction tuning datasets were construed for the English FinGPT model, we refer to the FinGPT repository.\footnote{\url{https://github.com/AI4Finance-Foundation/FinGPT}}

\subsubsection{Translation Approach}
\label{sec:translate}

To translate the instruction datasets to Dutch, we made use of \texttt{gpt-3.5-turbo-0125}\footnote{\url{https://platform.openai.com/docs/models/gpt-3-5-turbo}} of OpenAI. It should be noted that due to the extensive size of the datasets, only limited manual verification was performed on the translated texts. Consequently, some translated content might contain inaccuracies or non-factual details. We provide a script, inspired by \cite{vanroy2023language}, that uses OpenAI API services or open-source models to translate instruction tuning datasets.\footnote{\url{https://github.com/snoels/fingeit}} This approach can be adapted to translate datasets into any language supported by the translation model. This systematic but flexible translation approach has the potential to develop a diverse range of LLMs that surpass the boundaries of the English language. For more details on the translation process, see the Appendix \ref{appendix:translation-details}.

Given this context, it is crucial to acknowledge the potential drawbacks associated with automated translation. For instance, "translationese" effects might occur in the datasets, where the translated versions retain properties of the source text, such as word order and literal translation. This translation issue might lead to a biased advantage for non-Dutch models when evaluating on benchmarks. If these benchmarks were manually translated or originated in Dutch, it could limit the performance of non-Dutch models \cite{vanroy2023language}.

\subsubsection{Deduplication and Filtering}

Data duplication is a common issue in collected datasets, which can obstruct the efficiency and effectiveness of further model training. To mitigate this, we performed an exact deduplication process designed to identify and remove exact duplicate samples within our dataset.

\begin{table}[htbp]
\footnotesize
\centering
\begin{tabular}{p{4cm}p{2cm}}
\toprule
\textbf{Dataset} & \textbf{Instructions} \\ 
\midrule
Total English Instructions & 283,439 \\
Total Translated Instructions & 246,883 \\ 
Total Filtered and Deduplicated Instructions & 147,788 \\ 
\bottomrule
\end{tabular}
\caption{The count of instruction tuning samples throughout the data cleaning pipeline.}
\label{table:samples}
\end{table}

After deduplication, we proceeded with additional filtering. We applied a language identification filter, which excluded all non-Dutch documents post-translation. This method ensured that our dataset mainly contains Dutch language samples, which is necessary for developing a Dutch-specific financial LLM. To identify and filter out potential poor translations, an additional script was used. The language identification tool \textit{fasttext-language-identification}\footnote{\url{https://huggingface.co/facebook/fasttext-language-identification}} was used to confirm if the instruction tuning samples were correctly translated into Dutch. Additional checks were carried out to ensure the presence of 'Signifiers' and a minimum length of three tokens in the translations. This filtering process is mainly inspired by \cite{vanroy2023language}. Table \ref{table:samples} shows the number of instruction tuning samples throughout the pipeline. For a more comprehensive understanding of our deduplication and filtering pipeline, please refer to our repository.\footnote{\url{https://github.com/snoels/fingeit}}

\subsection{FinGEITje}
\label{subsec:fingeitje}

In this section, we present FinGEITje, our Dutch financial LLM. We describe the model architecture, training details, and future directions to enhance its capability in the Dutch financial sector.

For FinGEITje, we use \texttt{Mistral-7B-v0.1} \cite{jiang2023mistral} as our base model. This model is known for its superior performance, outperforming the LLaMA 13B model on numerous benchmarks. Most financial LLMs are built on smaller LLaMA models, but the robust Mistral BPE tokenizer, which efficiently segments numbers into single digits, makes it especially suitable for numerical tasks in the financial domain \cite{nie2024survey}.

\textbf{Domain-Specific Pretraining.} To implement language-specific adaptations, we use the foundation model \texttt{GEITje-7B}\footnote{\url{https://huggingface.co/Rijgersberg/GEITje-7B}}, a Dutch base model capable of handling sequences up to 8,192 tokens, and trained on over 10 billion tokens of Dutch text. This model serves as the linguistic backbone of FinGEITje.

\textbf{Financial Instruction Tuning.} Instruction tuning \cite{wei2021finetuned} aligns a model more closely with specific tasks while requiring less computational overhead compared to complete retraining. To efficiently handle instruction tuning, we employ FlashAttention \cite{dao2022flashattention}. Additionally, we use QLoRA \cite{dettmers2024qlora} to perform instruction fine-tuning, focusing on all linear layers, known to provide performance similar to full fine-tuning but with markedly reduced resource demands. We set a batch size of 8 and a learning rate of 2.0e-04. A 10\% warmup phase is applied, ensuring smooth adaptation to the new data. FinGEITje undergoes a single epoch of training on 2 NVIDIA A100 40GB GPUs. Detailed parameters and training configurations are available on the model card \texttt{snoels/FinGEITje-7B-sft}\footnote{\url{https://huggingface.co/snoels/FinGEITje-7B-sft}}.

\textbf{Preference Optimization.} While fine-tuning FinGEITje has provided a strong performance baseline, future iterations will incorporate preference optimization techniques, which are aimed at better aligning the model outputs with user expectations and improving response quality in interactive applications \cite{rafailov2024direct}.

\subsection{Dutch Financial Evaluation Benchmark}
\label{subsec:dutch-benchmark}

Evaluation benchmarks provide valuable information on the performance of a model, allowing us to identify key areas for improvement. Unfortunately, the availability of benchmarks for Dutch LLMs is disproportionally low \cite{vanroy2023language}. There are none existing within the financial context. Currently, evaluation datasets for Dutch LLM have been created by translating English benchmarks into Dutch \cite{lai2023okapi}, which lacks a genuine representation of the Dutch language.

To evaluate the performance of a Dutch financial LLM, a comprehensive and diverse benchmark is required that encompasses various key financial tasks. To this end, we introduce the first Dutch financial evaluation benchmark. It is designed based on our instruction tuning dataset construction method (see Section \ref{subsubsec:dataset-construction}). This benchmark covers critical tasks essential for effective assessment, such as Sentiment Analysis (SA), News Headline Classification (HC), Named Entity Recognition (NER), Relationship Extraction (RE), and Question Answering (QA).
\begin{table}[htbp]
\footnotesize
\centering
\begin{tabular}{p{2cm}ccc}
\toprule
\textbf{Dataset} & \textbf{Task} & \textbf{Test Samples} & \textbf{Evaluation Metrics} \\ 
\midrule
FPB \cite{malo2014good}, FiQA-SA \cite{maia201818}, TFNS \cite{magic2022twitter}, NWGI & SA & 5,788 & F1, Accuracy \\ 
NER (CLS) \cite{alvarado2015domain} & NER & 1959 & F1, Accuracy \\
Headline \cite{sinha2021impact} & HC & 9,094 & F1, Accuracy \\
FinRed \cite{sharma2022finred} & RE & 1,790 & F1, Accuracy \\
ConvFinQA \cite{chen2022convfinqa} & QA & 1,453 & Accuracy \\
\bottomrule
\end{tabular}
\caption{Summary of datasets used in the Dutch financial evaluation benchmark.}
\label{table:benchmark}
\end{table}

Table \ref{table:benchmark} provides an overview of the datasets included in the benchmark and the respective key evaluation metrics. Along with our instruction dataset, we release the Dutch financial evaluation benchmark and the corresponding code for automated model evaluation on Github\footnote{\url{https://github.com/snoels/fingeit}} and Huggingface\footnote{\url{https://huggingface.co/snoels/FinGEITje-7B-sft}}. By providing a comprehensive, task-specific benchmark, we aim to encourage more robust and meaningful comparisons and evaluations of Dutch financial LLMs.

\section{Experiments and Results}
\label{sec:experiments}

In this section, we discuss the experiments conducted to evaluate the performance of various LLMs, including our proposed FinGEITje, against our newly developed Dutch financial benchmark. We also discuss the methodology used for answer extraction and present detailed benchmarking results for both Dutch and English datasets.

\subsection{Experiments}

This subsection outlines the experimental setup and methodologies used to evaluate FinGEITje and other LLMs on our newly developed Dutch financial evaluation benchmark, providing insights into benchmarking and answer extraction methods.

\subsubsection{Benchmarking Methods}

We benchmark the performance of several LLMs, including FinGEITje, on our evaluation benchmark. We utilize two general Dutch instruction fine-tuned models: \texttt{GEITje-7B-ultra-sft}\footnote{\url{https://huggingface.co/BramVanroy/GEITje-7B-ultra-sft}} \cite{vanroy2023language} and \texttt{GEITje-7B-ultra}\footnote{\url{https://huggingface.co/BramVanroy/GEITje-7B-ultra}}. Furthermore, two English-based LlaMa-models \cite{touvron2023llama} designed explicitly for financial applications, FinMA \cite{xie2023pixiu} and FinGPT \cite{wang2023fingpt}, are selected. Furthermore, we evaluate a closed source, general-purpose GPT-3.5 model from OpenAI, \texttt{gpt-3.5-turbo-0125}\footnote{\url{https://platform.openai.com/docs/models/gpt-3-5-turbo}}. We assess the zero-shot performance of all the models against 500 randomly chosen test samples to decrease the computational cost.

\subsubsection{Answer Extraction}

Some LLMs without task-specific fine-tuning fail to generate answers pre-defined in the instructions \cite{xie2023pixiu,bhatia2024fintral}. Models might deviate from the task or fail to follow instructions completely. This discrepancy becomes problematic during model performance evaluation, particularly when outputs in a specific format are required, such as for QA and NER tasks. To address this, we propose applying an independent LLM as an evaluator tasked with extracting critical information from the generated responses to a task. We provide a script that uses the OpenAI API or open-source models to post-process the generated answers in the appropriate format as specified by the instruction.\footnote{\url{https://github.com/snoels/fingeit}} This method could be instrumental in conducting large-scale evaluations where human intervention is not viable. Details on the answer extraction process can be found in Appendix \ref{appendix:answer-extraction}. 

\subsection{Benchmark Results}

In this subsection, we present the comprehensive results of different LLMs against our Dutch-specific financial benchmark. Additionally, we evaluate the adaptability of FinGEITje by testing its performance against an English financial benchmark. This comparative analysis highlights the effectiveness of our Dutch financial LLM across different languages and tasks.

\subsubsection{Results on Dutch Data}

In this section, we present the results of different LLMs against our Dutch-specific financial benchmark. As shown in Table \ref{table:results-nl}, our proposed LLM, FinGEITje, outperforms the other models across all tasks, indicating the effectiveness of language-specific instruction tuning for domain-specific tasks. Interestingly, the Dutch-specific models, which have been fine-tuned for general Dutch tasks, struggle when applied to financially-focused benchmarks. Additionally, OpenAI's \texttt{gpt-3.5-turbo-0125} demonstrates strong zero-shot performance, highlighting its ability to perform tasks without detailed fine-tuning.

\begin{table*}[htbp]
\footnotesize
\centering
\begin{tabular}{@{}llcccccccccccccccc@{}}
\toprule
Dataset & Metrics & \multicolumn{2}{c}{\texttt{FinGEITje-7B-sft}} & \multicolumn{2}{c}{\texttt{geitje-7b-ultra}} & \multicolumn{2}{c}{\texttt{geitje-7b-chat}} & \multicolumn{2}{c}{\texttt{FinMA}} & \multicolumn{2}{c}{\texttt{FinGPT}} & \multicolumn{2}{c}{\texttt{gpt-3.5-turbo-0125}} \\ 
\cmidrule(l){3-14} 
 &  & raw & extracted & raw & extracted & raw & extracted & raw & extracted & raw & extracted & raw & extracted \\ \midrule
SA & Acc & \textbf{0.790} & \textbf{0.790} & 0.564 & 0.674 & 0.454 & 0.540 & 0.276 & 0.632 & 0.268 & 0.350  & \underline{0.742} & \underline{0.752} \\
& F1 & \textbf{0.790} & \textbf{0.790} & 0.555 & 0.662 & 0.466 & 0.520 & 0.136 & 0.645 & 0.113 & 0.276  & \underline{0.729} & \underline{0.740} \\
NER (CLS) & Acc & \textbf{0.840} & \textbf{0.912} & \underline{0.238} & 0.662 & \underline{0.238} & 0.670 & 0.236 & \underline{0.776} & 0.222 & 0.386  & \underline{0.238} & 0.690 \\
& F1 & \textbf{0.851} & \textbf{0.916} & \underline{0.092} & 0.684 & \underline{0.092} & 0.693 & 0.091 & \underline{0.792} & 0.089 & 0.357  & \underline{0.092} & 0.712 \\
HC & Acc & \textbf{0.920} & \textbf{0.920} & 0.082 & 0.064 & 0.314 & 0.298 & 0.000 & 0.670 & 0.000 & \underline{0.696} & \underline{0.606} & 0.640 \\
& F1 & \textbf{0.836} & \textbf{0.836} & 0.069 & 0.025 & 0.215 & 0.166 & 0.000 & 0.459 & 0.000 & 0.000 & \underline{0.466} & \underline{0.486} \\
RE & Acc & \textbf{0.569} & \textbf{0.569} & 0.000 & 0.111 & 0.000 & 0.123 & 0.003 & 0.045 & \underline{0.021} & \underline{0.259} & 0.000 & 0.157 \\
& F1 & \textbf{0.560} & \textbf{0.560} & 0.000 & 0.121 & 0.000 & 0.148 & 0.005 & 0.057 & \underline{0.015} & \underline{0.232} & 0.000 & 0.156 \\
QA & Acc & \textbf{0.324} & \textbf{0.324} & 0.016 & 0.036 & 0.056 & 0.056 & \underline{0.286} & \underline{0.294} & 0.011 & 0.008 & 0.196 & 0.244 \\
\bottomrule
\end{tabular}
\caption{Performance comparison across different models and datasets in Dutch. Best results are \textbf{bold}, second best are \underline{underlined}.}
\label{table:results-nl}
\end{table*}

\subsubsection{Results on English Data}

In order to test the adaptability of the Dutch financial language model, FinGEITje, we conducted a comparison of its performance against English financial evaluation data. The results are summarized in Table \ref{table:results-en}. Although English financial LLMs FinGPT and FinMA unsurprisingly perform best overall, FinGEITje, trained specifically on Dutch financial tasks, demonstrates notable competence. It emerges as the best model for NER (CLS), and manages to secure second highest performance for three tasks including SA, RE and QA. 

\begin{table*}[htbp]
\footnotesize
\centering
\begin{tabular}{@{}llcccccc@{}}
\toprule
Dataset & Metrics & \multicolumn{2}{c}{\texttt{FinGEITje-7B-sft}} & \multicolumn{2}{c}{\texttt{FinMA}} & \multicolumn{2}{c}{\texttt{FinGPT}} \\ 
\cmidrule(l){3-8} 
 &  & raw & extracted & raw & extracted & raw & extracted \\ \midrule
SA & Acc & \underline{0.684} & \underline{0.692} & \textbf{0.728} & \textbf{0.728} & 0.402 & 0.402 \\
& F1 & \underline{0.683} & \underline{0.693} & \textbf{0.729} & \textbf{0.729} & 0.247 & 0.248 \\
NER (CLS) & Acc & \textbf{0.806} & \textbf{0.880} & 0.402 & 0.484 & \underline{0.432} & \underline{0.492} \\
& F1 & \textbf{0.847} & \textbf{0.883} & 0.309 & 0.442 & \underline{0.365} & \underline{0.464} \\
HC & Acc & 0.366 & 0.638 & \textbf{0.884} & \textbf{0.884} & \underline{0.798} & \underline{0.798} \\
& F1 & 0.189 & \underline{0.290} & \textbf{0.797} & \textbf{0.797} & \underline{0.205} & 0.205 \\
RE & Acc & \underline{0.105} & \underline{0.309} & 0.004 & 0.088 & \textbf{0.589} & \textbf{0.614} \\
& F1 & \underline{0.115} & \underline{0.333} & 0.007 & 0.085 & \textbf{0.613} & \textbf{0.630} \\
QA & Acc & \underline{0.408} & \underline{0.420} & \textbf{0.466} & \textbf{0.495} & 0.004 & 0.004 \\
\bottomrule
\end{tabular}
\caption{Performance comparison across different models and datasets in English. Best results are \textbf{bold}, second best are \underline{underlined}.}
\label{table:results-en}
\end{table*}

\subsection{Discussion}

FinGEITje is the first Dutch financial language model, optimized with a specialized instruction tuning dataset. Our benchmarks show that FinGEITje outperforms general Dutch instruction-tuned models, general-purpose models, and even English finance-specific models in all Dutch financial tasks. Surprisingly, FinGEITje also performs well on English financial tasks.

Our answer extraction process effectively improves performance by converting raw outputs to the correct format. For instance, when FinGEITje received English instructions, its Dutch outputs were correctly translated back into English, demonstrating the process's reliability.

FinGEITje highlights the advantages of customizing language models for specific domains and languages. This approach is not only useful for Dutch but can also be applied to other languages. Our open-source instruction dataset translation method can help overcome the lack of datasets in non-English languages.

\section{Conclusion}
\label{sec:conclusion}

In this paper, we introduce FinGEITje, the first Dutch financial LLM fine-tuned on multiple financial tasks. We release a Dutch financial instruction tuning dataset comprising over more than 140,000 samples and an open-source data construction method that enables the creation of financial instruction datasets in various languages. In addition to the dataset, we provide the first Dutch financial evaluation benchmark. In addition, we present an open-source automated evaluation method, which employs an LLM as an independent evaluator, reducing the need for human intervention in LLM performance evaluation.

Our extensive experimental evaluation demonstrates FinGEITje's superior performance on 5 critical Dutch and English financial tasks. Trained with limited computational resources and maximum transparency, our approach sets a new standard for Dutch financial NLP.

Our open-source contributions aim to promote further research and innovation in financial natural language understanding, facilitating the development of useful and safe LLMs in finance.

\section{Future Work and Limitations}
\label{sec:futurework-limitations}

We reflect on possible future research directions and the current limitations of FinGEITje and similar financial LLMs, outlining potential enhancements.

\subsection{Future Work}

Our work on FinGEITje opens several promising avenues for future research. First, integrating advanced features such as vision, retrieval methods, and reinforcement learning approaches into the current model could potentially improve performance. Additionally, pre-training the Dutch foundational LLM on finance-specific texts may further improve domain-specific adaptation, and the application of domain-based filtering could also yield performance gains.

Moreover, expanding the scope of instruction tuning tasks is crucial for future improvements. This would allow the model to handle a wider range of financial tasks, thereby deepening its understanding of the financial domain. Challenging tasks, in particular, could benefit from increased instruction data volume.

Finally, the translation and instruction fine-tuning approach presented could be used as a framework for creating other domain-specific instruction tuning datasets and language-specific LLMs, thus improving the adaptability of LLMs across diverse realms.

\subsection{Limitations}

FinGEITje, alongside current state-of-the-art financial open-source LLMs, has several limitations. One significant constraint is the requirement for substantial computational resources, which poses a considerable barrier for researchers with limited computational capabilities. Additionally, since FinGEITje is specifically tailored for the finance domain, its performance may degrade when applied to tasks outside its specialized scope, potentially limiting its generalizability across diverse fields.

The model's capacity to handle complex reasoning tasks is also limited by the current architectural state of the 7B parameter models. Enhancing the reasoning abilities of the model remains a promising area for future research.

\section{Ethical Considerations}
\label{sec:ethics}

It is important to address ethical considerations when conducting research involving large language models in the financial domain. This section discusses key ethical aspects related to energy consumption, data privacy and copyright, bias and fairness, and the responsible use of FinGEITje.

\textbf{Energy Consumption.}
Training large language models requires significant computational resources, leading to considerable energy consumption and associated environmental impacts. We recognize the importance of sustainability in AI research and encourage ongoing efforts within the community to develop more energy-efficient algorithms and training techniques. Promoting sustainability is crucial for balancing technological advancement with ecological responsibility.

\textbf{Data Privacy and Copyright.}
Our datasets are constructed from publicly available financial texts, including news articles, reports, and social media posts. We have ensured compliance with copyright laws, and no proprietary or confidential data has been used in training FinGEITje. By utilizing publicly available data, we aim to respect the intellectual property rights of content creators while providing a valuable resource for financial natural language processing.

\textbf{Bias and Fairness.}
Language models can inadvertently learn and perpetuate biases present in their training data. FinGEITje is trained on a diverse range of financial documents to capture a broad spectrum of the financial domain. We acknowledge the possibility of residual biases and encourage users to critically assess the model's outputs, especially in contexts where fairness and impartiality are paramount. Continuous evaluation and feedback from the community are essential for addressing and mitigating potential biases.

\textbf{Responsible Use.}
FinGEITje is intended to support education and research. It is important that users employ the model ethically and responsibly, adhering to all relevant laws and regulations. We emphasize the significance of considering the societal impacts of deploying language models in financial contexts. The potential misuse of financial LLMs raises concerns about the dissemination of financial misinformation or unethical market influence. Users should avoid applications that could lead to unethical practices, such as market manipulation or the spread of misinformation. This underlines the need for rigorous mechanisms that ensure the responsible and ethical application of financial LLMs in real-world scenarios.

\begin{acks}
This research received funding from the Flemish government, through Flanders Innovation \& Entrepreneurship (VLAIO, project HBC.2020.2883) and from the Flemish government under the program “Onderzoeksprogramma Artificiële Intelligentie (AI) Vlaanderen”. 
\end{acks}

\bibliographystyle{ACM-Reference-Format}
\bibliography{main}

\appendix
\section{Instruction Dataset Translation Details}\label{appendix:translation-details}

All English instruction datasets were translated using \textit{gpt-3.5-turbo-0125}, with \textit{max\_tokens} set to \textit{1024} and \textit{temperature} to \textit{0.0}. For all tasks, the following system message and prompt were used:

\noindent

\mybox{%
\textbf{System message:} \\
You are a helpful assistant that translates English to \textcolor{blue}{<target\_language>} to the requirements that are given to you.
}

\noindent

\mybox{%
\textbf{Prompt:} \\
You are asked to translate a task’s instruction, optional input to the task, and the output of the task, from English into \textcolor{blue}{<target\_language>}. Here are the requirements that you should adhere to: 1. do not translate the identifiers ‘instruction: ‘, ‘input: ‘, and ‘response: ‘ but instead copy them to your output; 2. make sure that text is fluent to read and does not contain grammatical errors. Use standard \textcolor{blue}{<target\_language>} without regional bias; 3. translate the instruction and input text using informal, but standard, language; 4. make sure to avoid biases (such as gender bias, grammatical bias, social bias); 5. if the instruction is to correct grammar mistakes or spelling mistakes then you have to generate a similar mistake in the input in \textcolor{blue}{<target\_language>}, and then also generate a corrected output version in the output in \textcolor{blue}{<target\_language>}; 6. if the instruction is to translate text from one language to another, then you do not translate the text that needs to be translated in the instruction or the input, nor the translation in the output (just copy them as-is); 7. do not translate code fragments but copy them to your output. If there are English examples, variable names or definitions in code fragments, keep them in English; 8. maintain the format: the task consists of a task instruction (marked ‘instruction: ‘), optional input to the task (marked ‘input: ‘) and output for the task marked with ‘response: ‘. Now translate the following task with the requirements set out above. Do not provide an explanation and do not add anything else.
}

\par 

\section{Answer Extraction Prompt}
\label{appendix:answer-extraction}
\par
\raggedright 

The answers for all LLMs were extracted using \textit{gpt-3.5-turbo-0125} with \textit{max\_tokens} set to \textit{1024} and \textit{temperature} to \textit{0.0}. Here we provide the general system and user prompt; visit our GitHub repository for further information.

\noindent
\centering

\mybox{%
 \textbf{System message:} \\
 You are a helpful assistant specialized in extracting the label of a message: \\
 The possible labels are \textcolor{blue}{<scale\_string>}. If none of the labels apply reply with 'unknown'.
}

\mybox{%
 \textbf{Prompt:} \\
 \#\#\# Instruction: \\
 Determine the label of the message. \\
 Options: \textcolor{blue}{<scale\_string>}. \\
 No other options may be given. \\
 \#\#\# Input: \\
 \textcolor{blue}{<prediction>} \\
 \#\#\# Response:
}

\par
\raggedright 

The \textcolor{blue}{scale\_string} signifies the set of possible outputs for a given financial task in a certain language, and the \textcolor{blue}{prediction} represents the raw prediction of an LLM for a specified financial instruction task.

\end{document}